\documentclass[10pt]{article}

\usepackage[english]{babel}
\usepackage[a4paper,top=2cm,bottom=2cm,left=3cm,right=3cm,marginparwidth=1.75cm]{geometry}
\usepackage[backend=biber,style=nature]{biblatex}
\usepackage{amsmath, amssymb, amsfonts, amsthm}
\usepackage{mathrsfs}
\usepackage{cancel}
\usepackage{xfrac}
\usepackage{graphicx}
\usepackage{caption}
\usepackage{subcaption}
\usepackage[colorlinks=true, allcolors=blue]{hyperref}
\usepackage[auth-sc, affil-it]{authblk}
\usepackage{parskip}
\usepackage{textcomp}
\usepackage{color,soul}
\usepackage{xcolor}
\usepackage{multirow}
\usepackage{booktabs}

\usepackage[title]{appendix}
\usepackage{algorithm}
\usepackage{algorithmicx}
\usepackage{algpseudocode}
\usepackage{listings}
\usepackage{manyfoot}
\usepackage{longtable}
\usepackage{listings}
\usepackage{xcolor}
\usepackage{csquotes}
\usepackage{tabularx}
\usepackage{bm}
\usepackage[table]{xcolor} 

\title{Annotated digital image correlation displacement fields from fatigue crack growth experiments}

\author[1,*]{David Melching}
\author[1]{Ferdinand Dömling}
\author[1]{Florian Paysan}
\author[1]{Erik Schultheis}
\author[1]{Eric Dietrich}
\author[1]{Eric Breitbarth}

\affil[1]{Institute for Frontier Materials on Earth and in Space, German Aerospace Center (DLR), Linder Hoehe, Cologne, 51147, Germany}
\affil[*]{Corresponding author: David.Melching@dlr.de}

\date{February 11, 2026}

\begin{document}
\maketitle

\begin{abstract}
We present a curated dataset of planar displacement fields from eight fatigue crack growth experiments obtained via full-field digital image correlation (DIC). The dataset covers multiple aerospace-grade aluminium alloys, specimen geometries, material orientations, and load configurations, providing a diverse experimental basis for data-driven fracture mechanics research. 

Crack tip locations are consistently annotated using an iterative correction procedure applied to all measurements, and fracture mechanical descriptors like stress-intensity factors are provided as additional labels. The dataset comprises 8,794 unique experimentally observed displacement fields and a total of 70,352 supervised samples generated through standardized interpolation and augmentation.

DIC data is provided as uniformly interpolated displacement grids at three standardized resolutions (\(28 \times 28\), \(64 \times 64\), and \(128 \times 128\) pixels), each available in three dataset sizes to support scalable use cases ranging from educational applications to high-capacity model development. Accompanying metadata and a Python interface facilitate filtering, loading, and integration into reproducible machine learning and fracture mechanics workflows.
\end{abstract}

\section*{Background \& Summary}
Fatigue crack growth (FCG) experiments are fundamental in materials science and structural engineering because they inform lifetime prediction and safe design. Conventional crack length measurements, such as the potential drop technique, provide robust estimates but lack the spatial detail needed to analyse local crack growth mechanisms. Analytical solutions and numerical simulations can estimate the stress state near the crack-tip, but they rely on idealised assumptions and therefore provide only indirect insight into local crack-tip behaviour. Optical methods such as digital image correlation (DIC) bridge this gap. DIC delivers planar full-field displacement maps throughout crack propagation, supporting quantitative analysis of near-crack-tip fracture mechanics as well as global full-field analysis of boundary conditions and introduced forces. Robust crack tip localization in DIC remains challenging due to measurement noise, limited spatial resolution, and the subtle gradients that characterize the elastoplastic fields near the tip.

In parallel, machine learning (ML) and deep learning (DL) approaches have gained increasing attention in fracture mechanics. Convolutional neural networks (CNNs) have been used to predict full-field stress distributions and stress concentrations in cracked or damaged structures \cite{BOLANDI2022103240}. Graph-based and transfer-learning frameworks have been developed to emulate crack propagation and stress intensity factors in brittle fracture problems \cite{PERERA2023104639}. Interpretable machine learning approaches like symbolic regression have also been proposed to construct surrogate models for stress intensity factors by learning corrections to classical analytical solutions \cite{MERRELL2024110432}. More recently, physics-enhanced deep learning models have been introduced to directly predict stress intensity factors for complex crack configurations by combining fracture-mechanics features with CNNs \cite{ZHAO2025110720}.
However, the training data underpinning these approaches are exclusively derived from numerical simulations, often based on idealized geometries, simplified material behavior, and noise-free fields. While such synthetic datasets are well suited for proof-of-concept studies, they only partially reflect the complexities of experimental measurements, including optical noise, spatial resolution limits, and deviations from ideal boundary conditions.
More recently, deep-learning-based approaches have been proposed for crack path and crack tip detection in experimental settings involving DIC data \cite{Strohmann21, Melching22}, as well as physics-informed iterative correction schemes achieving sub-pixel crack tip accuracy \cite{Melching24}. 
Although these studies clearly demonstrate the potential of data-driven and hybrid methods for crack-tip analysis in DIC, their broader comparison, reproducibility, and systematic benchmarking remain constrained by the absence of standardized, openly available, and consistently annotated experimental datasets based on full-field measurements. Existing dataset-focused contributions in the field, such as Mechanical MNIST Crack Path \cite{Mohammadzadeh2022} or large-scale fracture simulation repositories \cite{Hill2025}, are likewise limited to synthetic data, underscoring the need for experimentally grounded benchmarks that capture the real-world challenges of DIC-based fatigue crack growth analysis.

This work addresses this need by assembling a curated and annotated dataset called \textit{CrackMNIST}, which consists of full-field displacement data acquired during eight FCG experiments with a fully automated robotic test stand and a commercial Zeiss Aramis system \cite{Paysan2023,Strohmann24}. Parts of the underlying experimental data and analysis workflows have previously been used to investigate automated fatigue testing \cite{Strohmann24}, crack growth monitoring \cite{Strohmann21, Melching22} and fracture mechanical evaluations \cite{Paysan2023, Kalina2023, Ricoeur2024, Paysan2025, Schoene2025}. The present dataset consolidates these data into a unified, standardized, and openly accessible resource with consistent preprocessing and annotation across all experiments. Ground-truth crack tip positions are provided via an iterative correction procedure applied consistently across all experiments \cite{Melching24}. In addition to crack tip coordinates, the dataset includes fracture-mechanics-related quantities derived from the near-tip displacement fields. In particular, stress intensity factors are provided as annotations, enabling supervised learning approaches that map DIC displacement fields directly to physically meaningful fracture descriptors. To support a broad range of use cases, from rapid prototyping to high-fidelity modelling, the dataset features interpolated DIC displacement fields at various resolutions ranging from $28 \times 28$ to $128 \times 128$ pixels, all representing the same physical region of roughly $40-60\,\mathrm{mm} \times 40-60\,\mathrm{mm}$ on the specimen. For each resolution, three dataset sizes (S, M, and L) are provided, enabling flexible scalability for algorithm development and offering lightweight subsets tailored to educational applications.

To summarize, the published per-resolution-datasets provide a total of 70,352 supervised samples derived from 8,794 unique experimentally observed displacement fields. 

Key characteristics of the dataset include:

\textbf{Diversity} -- Eight experiments spanning AA2024, AA7475, and AA7010 alloys and covering multiple specimen geometries, thicknesses, fracture toughness orientations, and load ratios.

\textbf{Scalability} -- The DIC displacement fields are interpolated to regular grids at several pixel resoultions, $28 \times 28$, $64 \times 64$, and $128 \times 128$ pixels. Each resolution comes in three dataset sizes (S, M, L) supporting both lightweight exploration at a low entry-barrier as well as high-capacity, high-fidelity model training.

\textbf{Reuse potential} -- The dataset supports supervised learning tasks including crack tip localization and the direct regression of fracture-mechanics descriptors such as stress intensity factors from displacement fields, as well as studies on resolution dependence, data efficiency, uncertainty quantification.

\textbf{Education} -- The consistent structure, varied experimental cases, and multiple resolutions, particularly the $28 \times 28$ variant, make the dataset suitable for teaching modern machine-learning-based computer vision techniques using experimentally acquired fracture mechanics data.

\section*{Methods}
\begin{figure}
    \centering
    \includegraphics[width=\textwidth]{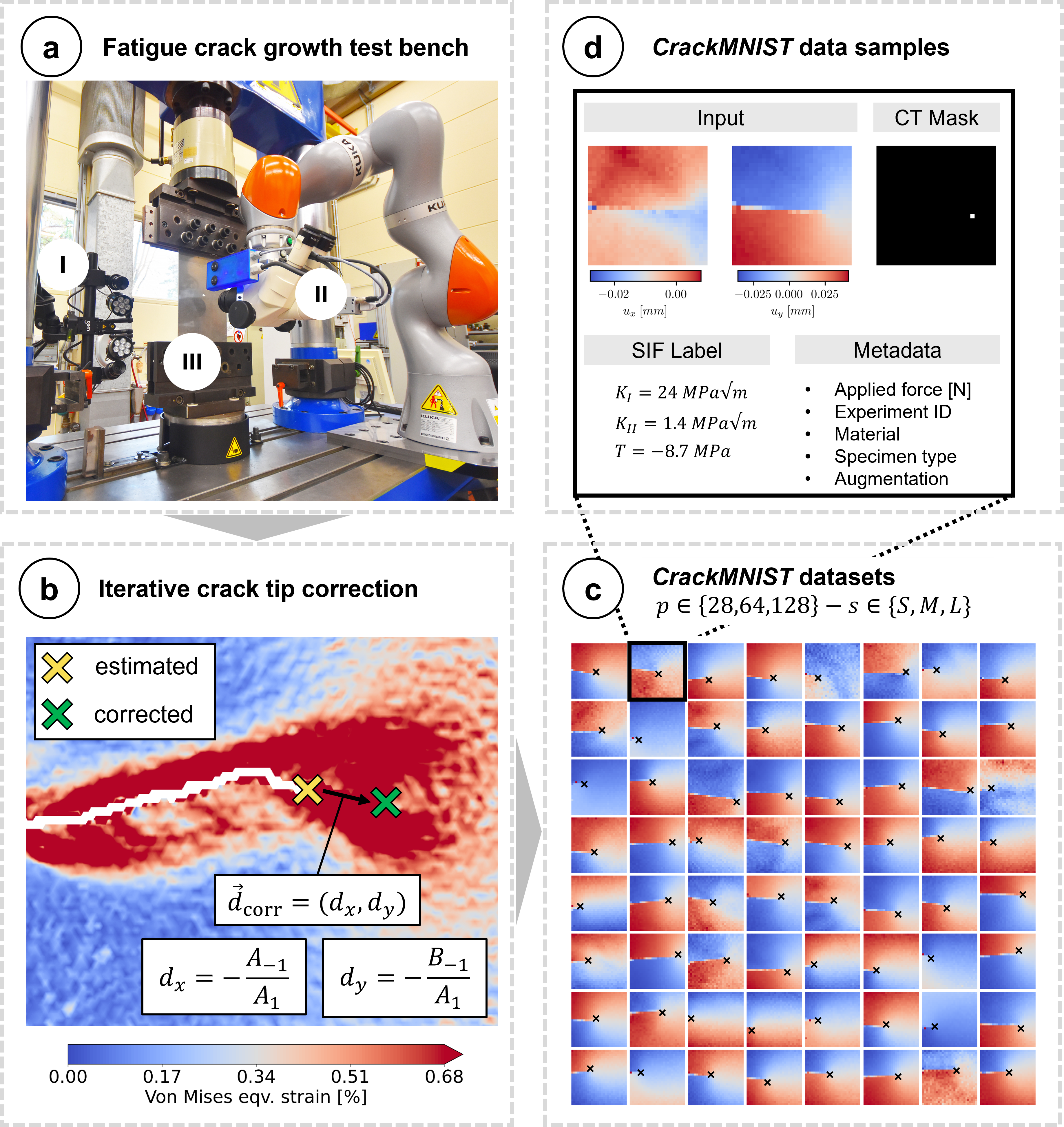}
    \caption{Overview of the \textit{CrackMNIST} dataset creation workflow:
    \textbf{(a)} Experimental setup for fatigue crack growth testing \cite{Strohmann24} with \textbf{I}: full-field 3D DIC by a commercial Zeiss Aramis system, \textbf{II}: high-resolution DIC option using a KUKA collaborative robot with a global shutter camera (not used in this dataset release), and \textbf{III}: material specimen (here: MT160). 
    \textbf{(b)} Crack tip annotation process showing initial estimate (yellow cross), refined position after iterative correction (green cross), and the symbolic regression correction formulas applied to Williams-series coefficients. 
    \textbf{(c)} Examples of processed displacement field samples at $28\times28$ pixel resolution with crack tip labels (black crosses) as provided in the \textit{CrackMNIST} dataset. The dataset is available at pixel resolutions, $p \in \{28, 64, 128\}$, and with various sizes, $s\in \{S,M,L\}$.
    \textbf{(d)} Data samples consist of input displacement field $(u_x, u_y)$ with corresponding target crack tip masks and stress intensity factors. For each sample metadata such as the applied external force and specimen and material type are provided.}
    \label{fig:crackmnist_graphical_abstract}
\end{figure}

The creation of \textit{CrackMNIST} followed a structured workflow from controlled fatigue crack growth experiments to standardized machine-learning-ready datasets (Fig.~\ref{fig:crackmnist_graphical_abstract}). Full-field 3D digital image correlation (DIC) measurements were acquired on aerospace-grade aluminium alloy specimens. Crack tip positions were then annotated with a high-accuracy iterative correction procedure based on fitting the experimental full-field displacements to the theoretical Williams field leading to estimated crack tip accuracies of 0.02 mm, cf. Section "Crack Tip Annotation". Mode-I and mode-II stress intensity factors, $K_I, K_{II}$, together with the T-stress, $T$, were calculated by fitting the Williams series expansion using \textit{CrackPy} \cite{Crackpy}. Finally, both displacement fields were randomly augmented and interpolated to multiple pixel resolutions.

\textbf{Experimental campaign.}
Fatigue crack growth experiments were conducted in accordance with ASTM~E647-15\cite{ASTM_E64715} using a servo-hydraulic uniaxial testing machine. A Zeiss Aramis 3D DIC system observed the rear side while a robotic high-resolution 2D (HR-DIC) system was used to track the front side crack-tip region for detailed investigation of the plastic zone. For the \textit{CrackMNIST} dataset, only the global 3D DIC data are considered, as the HR-DIC data is restricted to a highly localized field of view.
 
The conducted experiments are summarised in Table~\ref{tab:experiments}, covering aluminium alloys AA2024, AA7475, and AA7010 in rolled ($r$) or forged ($f$) condition, with MT160 and CT75 specimen geometries, thicknesses between 2--12\,mm, and multiple fracture orientations (LT, TL, SL45$^\circ$ following ASTM nomenclature for material orientation \cite{ASTM_E1823}) and load ratios $R$. For this test campaign, the maximum load $F_{\max}$ was kept constant for all $R$, while the minimum load $F_{\min}$ was adjusted to achieve the prescribed $R$-ratio. All tests were performed under sinusoidal, constant-amplitude loading with a frequency of 20 Hz.

\begin{longtable}{@{}llllll@{}}
\toprule
Experiment & Material & Specimen Type & Thickness [mm] & Orientation & R \\ \midrule
\endfirsthead
\toprule
Experiment & Material & Specimen Type & Thickness [mm] & Orientation & R \\ \midrule
\endhead
MT160\_2024\_LT\_1  & AA2024-T3$^r$ & MT160 & 2  & LT    & 0.1 \\
MT160\_2024\_LT\_2  & AA2024-T3$^r$ & MT160 & 2  & LT    & 0.3 \\
MT160\_2024\_LT\_3  & AA2024-T3$^r$ & MT160 & 2  & LT    & 0.5 \\
MT160\_2024\_TL\_1  & AA2024-T3$^r$ & MT160 & 2  & TL    & 0.1 \\
MT160\_2024\_TL\_2  & AA2024-T3$^r$ & MT160 & 2  & TL    & 0.3 \\
MT160\_7475\_LT\_1  & AA7475-T761$^r$ & MT160 & 4  & LT    & 0.1 \\
MT160\_7475\_TL\_1  & AA7475-T761$^r$ & MT160 & 4  & TL    & 0.3 \\
CT75\_7010\_SL45\_1 & AA7010-T7451$^f$ & CT75  & 12 & SL45$^\circ$ & 0.1 \\ \bottomrule
\noindent $^r$Rolled, $^f$Forged.\\
\caption{Experiments included in \textit{CrackMNIST} consist of different materials, specimen types, specimen thicknesses, fracture toughness orientations, stress ratios $R$.}
\label{tab:experiments}
\end{longtable}

\textbf{Digital image correlation system.} The full-field 3D DIC system (Zeiss Aramis 12M) comprised two 12\,MP cameras (4\,619\,$\times$\,2\,598\,pixels) mounted with a slider distance of 98\,mm and a mutual viewing angle of $25^\circ$ on a stereo base.
The measurement volume was configured to $200\times150\times21$\,mm$^3$ using 50\,mm lenses, with the camera pair positioned 525\,mm  from the specimen surface. Uniform illumination was provided by two 20\,W, blue-light (455-475\,nm). The specimen backside was prepared with a stochastic, high-contrast vanish spray pattern, suitable for DIC feature tracking. Displacement evaluation was performed with a facet size of $19\times19$\,pixels (0.86\,$\times$\,0.86\,mm$^2$) and a facet spacing of 16\,pixels (0.59\,$\times$\,0.59\,mm$^2$).

\textbf{Data acquisition.}
During testing, a direct current potential drop (DCPD) system monitored crack length in real time and triggered image acquisition. After each crack growth increment of $\Delta a_{\mathrm{step}}=0.5$\,mm, the DIC systems recorded data at at least three discrete loads corresponding to the minimum $F_{\min}$, an intermediate $(F_{\min}+F_{\max})/2$, and the maximum $F_{\max}$ load level. This automation enabled the acquisition of spatially and temporally resolved displacement fields throughout each experiment.

\textbf{Crack tip annotation.}
Crack tip positions were annotated in physical coordinates (mm) on the raw DIC data using a three-stage \texttt{CrackPy} \cite{Crackpy} workflow. First, the displacements, strains and derived stresses were processed for each specimen side, with the left side mirrored to ensure a consistent coordinate system. An initial crack path, angle, and tip estimate $(x_0, y_0, \theta)$ was obtained using a line-intercept detector operating on a regular grid of probe lines with a spacing of $0.1$\,mm in both the $x$- and $y$-directions. To ensure a physically correct crack tip position, this rough estimate was then refined by iterative fitting of the Williams series to the measured fields, using the correction formulas discovered by symbolic regression in \cite{Melching24}:
\begin{equation}
    d_x = -\frac{A_{-1}}{A_1}, \quad d_y = -\frac{B_{-1}}{A_1}, \quad \varepsilon = \sqrt{A_{-1}^2 + B_{-1}^2}, \quad \delta = \sqrt{d_x^2 + d_y^2}.
\end{equation}
The iteration was terminated once the correction magnitude $\delta$ fell below $5 \cdot 10^{-3}$ mm. Finally, the correction with the smallest error, $\varepsilon$, was selected. Analysis of the convergence behavior reported in \cite{Melching24} indicates geometrically shrinking corrections with a contraction factor $q \approx 0.6-0.8$. Under this assumption, the residual numerical uncertainty due to early stopping is bounded by $\delta/(1-q)$, corresponding to an expected termination error of approximately $0.02$\,mm for the chosen tolerance. This numerical contribution is small compared to the intrinsic model and measurement uncertainty of the Williams-fit-based correction reported in \cite{Melching24}.
All annotations were performed on the native DIC facet grid and subsequently mapped to the standardized interpolated image grids (28$\times$28, 64$\times$64, 128$\times$128) by applying the same affine transformation used for displacement interpolation. The labels are provided as binary segmentation masks with the tip pixel marked positive. 

\textbf{SIF annotation.}
Stress intensity factors (SIFs) $K_{\mathrm{I}}$, $K_{\mathrm{II}}$, and the non-singular $T$-stress were annotated for each data point by fitting the Williams series to the measured DIC displacement fields using the \texttt{CrackPy} fracture analysis framework \cite{Crackpy}. 
Starting from the refined crack-tip position and crack angle obtained during crack tip annotation, the fracture analysis workflow performed an iterative least-squares optimization of the Williams expansion over a prescribed set of terms and radial fitting bounds. 
For each data point, the pipeline evaluated the analytical displacement fields for the selected Williams modes and minimized the residuals to determine the coefficients $A_n$ and $B_n$. 
The mode-I and mode-II SIFs and T-stress were then computed directly from the coefficients as
\begin{equation}
    K_{\mathrm{I}} = \sqrt{2\pi}\,A_1, \qquad
    K_{\mathrm{II}} = - \sqrt{2\pi}\,B_1, \qquad
    T = 4 A_2.
\end{equation}

\textbf{Metadata annotation.}
Each image and its corresponding crack tip mask or SIF target are accompanied by a set of metadata that extends the static, experiment-level descriptors (Table~\ref{tab:experiments}) with per-step information, such as applied force. This metadata is intended to enable custom training and evaluation strategies as well as physically meaningful data selection. An overview of metadata fields available for query is provided in Table~\ref{tab:metadata}.

\begin{table}[ht]
\centering
\setlength{\tabcolsep}{4pt}
\renewcommand{\arraystretch}{1.15}
\begin{tabular}{
p{0.16\textwidth}  
p{0.14\textwidth}  
p{0.16\textwidth}  
p{0.46\textwidth}  
}
\toprule
\textbf{Descriptor} & \textbf{Type} & \textbf{Values} & \textbf{Explanation} \\
\midrule
Experiment  &  Categorical  & see Tab.~\ref{tab:experiments}  & Unique identifier of the experiment   \\
Material  &  Categorical  & see Tab.~\ref{tab:experiments}  & Alloy designation of the tested material   \\
Specimen Type  &  Categorical  & see Tab.~\ref{tab:experiments}  &  Geometrical specimen configuration\cite{ASTM_E1823}\\
Thickness  &  Categorical  & see Tab.~\ref{tab:experiments}  & Nominal specimen thickness\\
Orientation  &  Categorical  & see Tab.~\ref{tab:experiments}  & Crack plane/ Mode I loading direction relative to material rolling directions\\
R  &  Categorical  & see Tab.~\ref{tab:experiments}  & Nominal load ratio ($R=F_{\min}/F_{\max}$) of the fatigue cycle\\
Side  &  Categorical  &  left, right  &  Imaged side relative to \textit{pre-crack} notch geometry; distinguishes in left and right cracks for symmetric specimens (MT160).\\
Force  &  Continuous  & $[200, 30000]$\ $N$  & Applied uniaxial load during image acquisition. \\
\bottomrule
\end{tabular}
\caption{Overview of the  provided metadata fields in \textit{CrackMNIST}. Each Descriptor is accessible via the data loader (see \emph{Usage Notes}). }
\label{tab:metadata}
\end{table}

\begin{figure}[ht]
    \centering
    \includegraphics[width=0.9\linewidth]{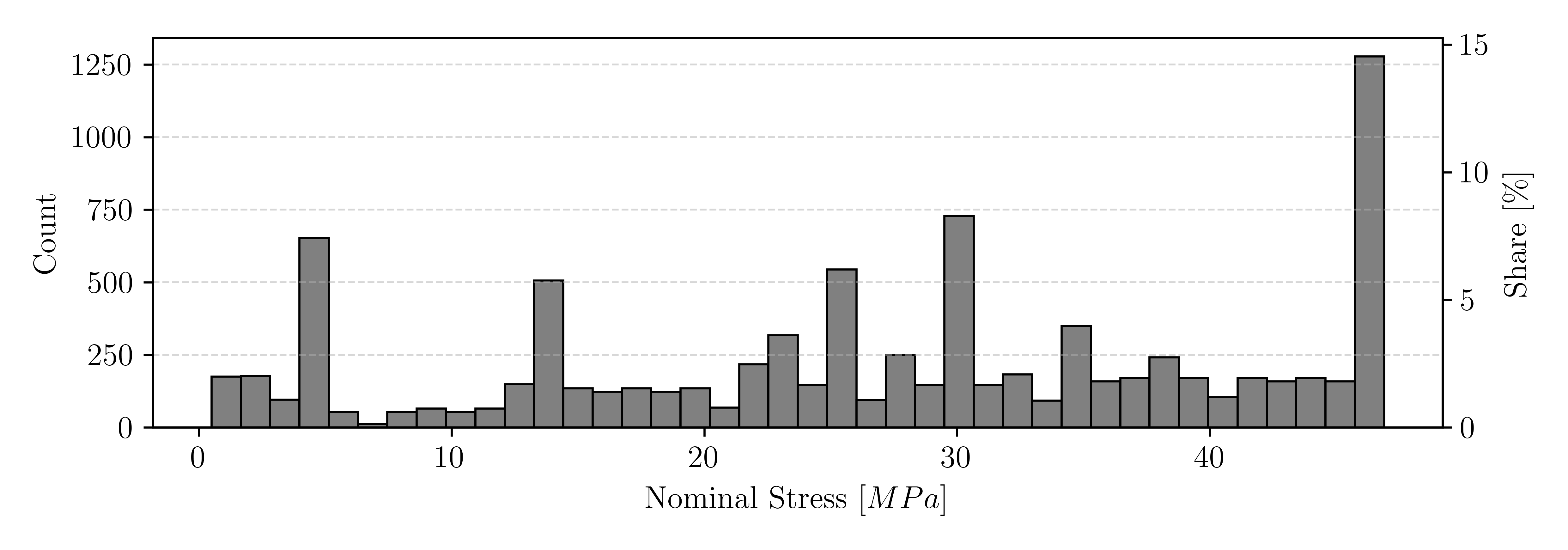}
    \caption{Distribution of samples across applied normalized loads. Counts are based on non-augmented data.}
    \label{fig:force_distribution}
\end{figure}

Table~\ref{tab:metadata_distribution} summarizes the distribution of the 8,794 unique experimentally observed displacement fields across the available metadata descriptors underlying the \textit{CrackMNIST} dataset. Figure~\ref{fig:force_distribution} shows the distribution of the applied nominal stresses, calculated as applied external force divided by the product of specimen thickness and width, using 40 bins. Of all nominal stresses, 43\% fall into bins corresponding to $\sigma_{\min}$, $(\sigma_{\min}+\sigma_{\max})/2$ and $\sigma_{\max}$ in equal proportions. The majority of 57\% is accounted for by further intermediate loads between $\sigma_{\min}$ and $\sigma_{\max}$.

\begin{table}[ht]
\centering
\setlength{\tabcolsep}{4pt}
\begin{tabularx}{\textwidth}{lXXX}
\hline
\textbf{Descriptor} & \textbf{Value} & \textbf{Count} & \textbf{Share [\%]} \\
\hline
experiment & MT160\_2024\_TL\_2 & 1486 & 16.90 \\
 & MT160\_7475\_LT\_1 & 1454 & 16.53 \\
 & MT160\_2024\_LT\_2 & 1276 & 14.51 \\
 & MT160\_2024\_LT\_1 & 1276 & 14.51 \\
 & MT160\_2024\_LT\_3 & 1232 & 14.01 \\
 & MT160\_2024\_TL\_1 & 743 & 8.45 \\
 & MT160\_7475\_TL\_1 & 716 & 8.14 \\
 & CT75\_7010\_SL45\_1 & 611 & 6.95 \\
\hline
material & AA2024 (rolled) & 6013 & 68.38 \\
 & AA7475 (rolled) & 2170 & 24.68 \\
 & AA7010 (forged) & 611 & 6.95 \\
\hline
specimen\_type & MT160 & 8183 & 93.05 \\
 & CT75 & 611 & 6.95 \\
\hline
thickness\_mm & 2 & 6013 & 68.38 \\
 & 4 & 2170 & 24.68 \\
 & 12 & 611 & 6.95 \\
\hline
orientation & LT & 5238 & 59.56 \\
 & TL & 2945 & 33.49 \\
 & SL45° & 611 & 6.95 \\
\hline
R & 0.1 & 4084 & 46.44 \\
 & 0.3 & 3478 & 39.55 \\
 & 0.5 & 1232 & 14.01 \\
\hline
side & right & 4721 & 53.68 \\
 & left & 4073 & 46.32 \\
\hline
\end{tabularx}
\caption{Distribution of samples across categorical metadata descriptors. Counts are based on non-augmented data.}
\label{tab:metadata_distribution}
\end{table}

\begin{figure}[ht]
    \centering
    \includegraphics[width=0.65\linewidth]{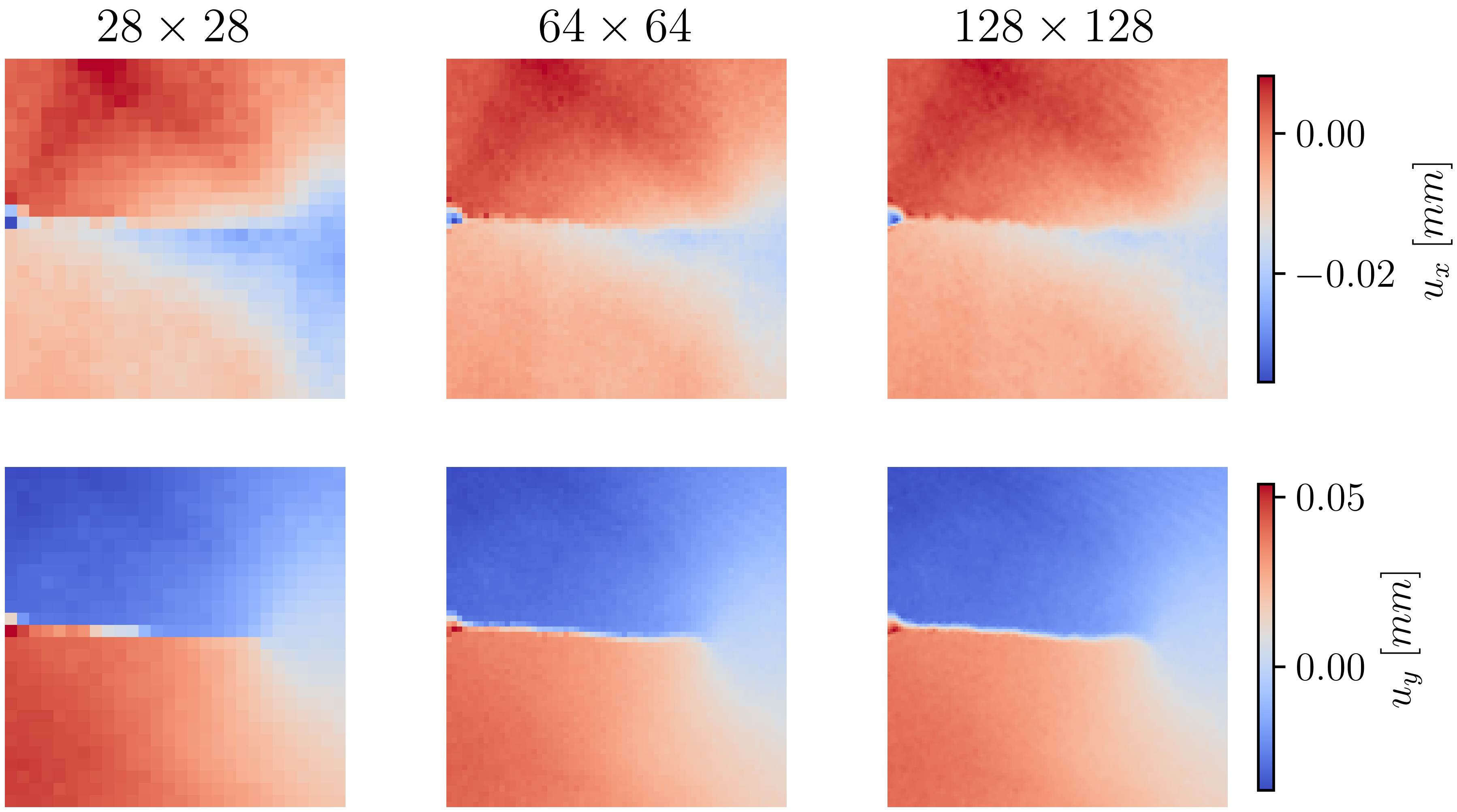}
    \caption{\textit{CrackMNIST} DIC data sample interpolated to different resolutions. Top: $x$-displacement in $mm$. Bottom: $y$-displacement in $mm$. Columns show the same sample at various resolutions.}
    \label{fig:resolutions}
\end{figure}

\textbf{Resolutions and interpolation.}
All annotations and displacement fields were generated on the raw DIC facet grid obtained from the Zeiss Aramis system. For machine learning applications and standardization across experiments, the raw fields were interpolated onto equidistant square grids of size $P \times P$, with $P \in \{28, 64, 128\}$. The resulting grids represent a physical field of view of approximately $40 \times 40\,\mathrm{mm}^2$ and $60 \times 60\,\mathrm{mm}^2$ for the CT- and MT-specimen, respectively. Interpolation was performed using bilinear mapping in the specimen’s physical coordinate frame, ensuring that both displacement components $(u_x, u_y)$ and crack tip positions were transformed consistently. The lower resolutions provide a lightweight \textit{MNIST}-style dataset suitable for educational purposes and rapid prototyping, while higher resolutions preserve fine-scale displacement gradients essential for high-fidelity crack tip detection. It is important to note that interpolation changes only the sampling density of the measured fields and does not increase the underlying measurement accuracy set by the original DIC facet spacing.
Figure~\ref{fig:resolutions} shows a DIC sample datapoint at the different resolutions provided by \textit{CrackMNIST}.

\begin{figure}
    \centering
    \includegraphics[width=0.75\linewidth]{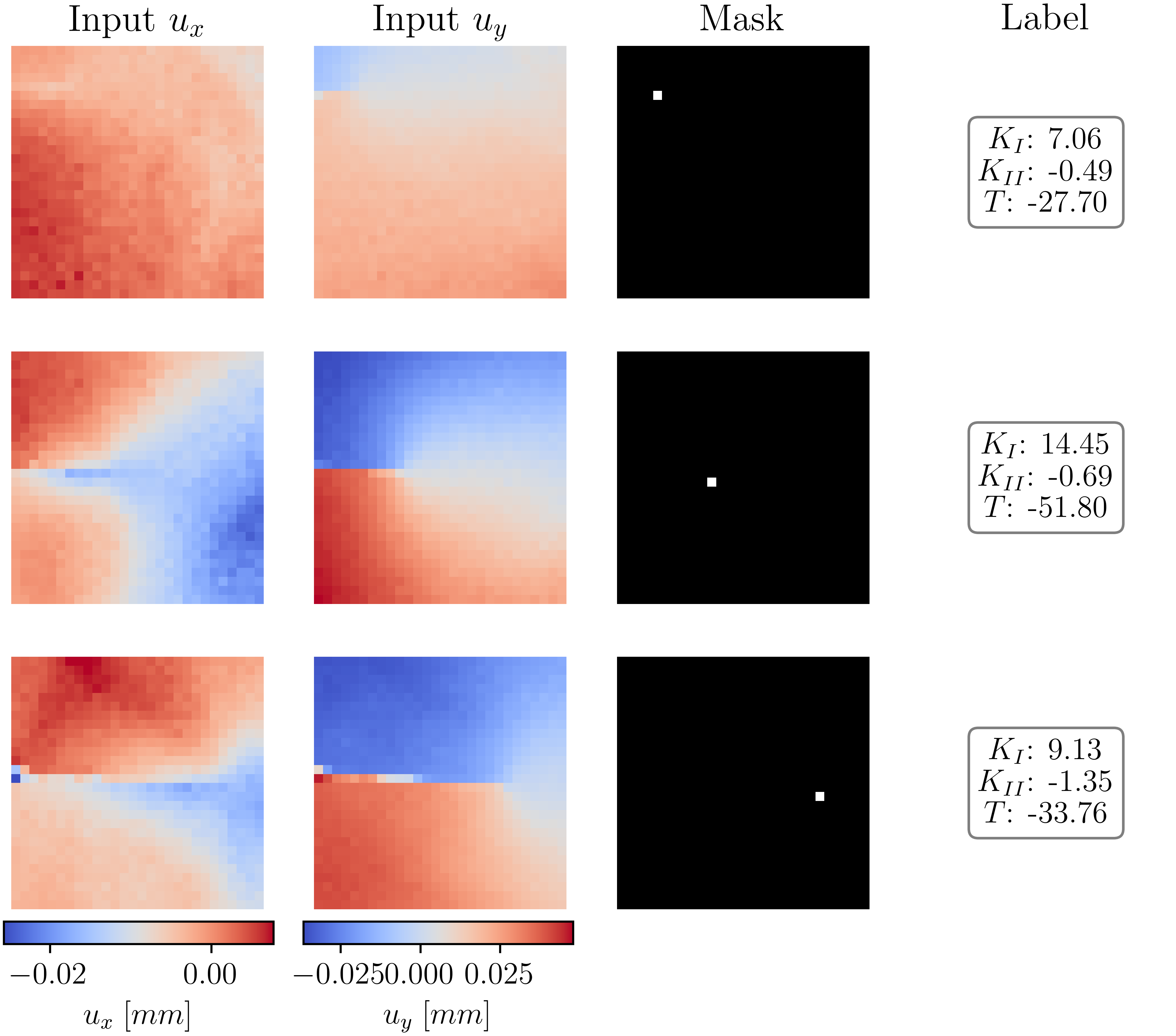}
    \caption{Three random \textit{CrackMNIST} samples at a resolution of $28 \times 28$ pixels. Each input consists of two channels representing the in-plane displacement components, $u_x$ and $u_y$, while the corresponding binary segmentation masks indicate the crack tip location, shown as white pixels. Additional annotations of the stress-intensity factors, $K_I, K_{II}$, both in $\mathrm{MPa}\sqrt{m}$, and the $T$-stress in $\mathrm{MPa}$ are given.}
    \label{fig:crackmnist_samples}
\end{figure}

\textbf{Processing and augmentation.}
In general, all experiments with a middle tension (MT) specimen contain a left- and right-hand side. For consistency, the left-hand side is rotated to the right such that the crack always grows in the same direction.
For each original DIC data point, 8 augmented variants are generated by applying random shifts ($x$ up to 10\,mm, $y \pm 10$ mm), rotations ($\pm$\,10$^\circ$), and a 50\% chance of vertical flips. The augmented crack tip position is transformed accordingly, and samples producing NaNs are discarded and re-drawn.
Augmentations are first validated at 256\,px, then applied identically to produce aligned versions at 28, 64, 128\,px. This ensures all resolutions contain \emph{exactly the same physical samples}, differing only in interpolation grid spacing. Labels are binary masks with value 255 (white/foreground) at the crack tip and 0 (black/background) elsewhere. For higher resolutions, a single crack tip pixel occupies a negligible fraction of the image, making it difficult to detect. To mitigate this, the crack tip label is dilated by one pixel in all directions, creating a small cluster of labelled pixels around the tip to improve visibility and learning stability. Therefore, to keep the crack-tip-to-background ratio relatively constant the crack tip masks are enlarged to 3$\times$3 and 5$\times$5 for the resolutions 64$\times$64 and 128$\times$128, respectively.
Inputs contain two channels: The planar displacements in $x$-direction, $u_x$, and in $y$-direction, $u_y$.
Figure \ref{fig:crackmnist_samples} displays the displacement fields and crack tip segmentation masks of \textit{CrackMNIST} data samples with various crack lengths.

\textbf{Dataset sizes.} To support scalable benchmarking and usability without large computational resources (e.g. educational demonstration and teaching), the augmented dataset comprising 70,352 unique samples is released at each resolution in three dataset sizes (S, M, L). Fixed splits are used and are identical across resolutions.
Table~\ref{tab:dataset_splits} summarizes the dataset splits for each size (S, M, L), listing the corresponding experiments (with left/right indicating the side of the specimen) used for training, validation, and test sets.

\begin{table}[ht]
\centering
\begin{tabularx}{\textwidth}{lXXX}
\toprule
\textbf{Size} & \textbf{Train} & \textbf{Validation} & \textbf{Test} \\
\midrule
S & 
MT160\_2024\_LT\_1\_right, MT160\_2024\_LT\_3\_right &
MT160\_2024\_TL\_2\_left &
MT160\_2024\_TL\_1\_right \\
\hline
M &
MT160\_2024\_LT\_3\_right, MT160\_2024\_LT\_3\_left, MT160\_2024\_TL\_2\_right, MT160\_7475\_LT\_1\_right &
MT160\_2024\_TL\_2\_left, MT160\_7475\_LT\_1\_left &
MT160\_2024\_TL\_1\_right, MT160\_7475\_TL\_1\_left \\
\hline
L &
MT160\_2024\_LT\_1\_right, MT160\_2024\_LT\_1\_left, MT160\_2024\_LT\_2\_right, MT160\_2024\_LT\_2\_left, MT160\_2024\_LT\_3\_right, MT160\_2024\_LT\_3\_left, MT160\_2024\_TL\_2\_right, MT160\_7475\_LT\_1\_right &
MT160\_2024\_TL\_2\_left, MT160\_7475\_LT\_1\_left &
MT160\_2024\_TL\_1\_right, CT75\_7010\_SL45\_1\_right, MT160\_7475\_TL\_1\_left \\
\bottomrule
\end{tabularx}
\caption{Experimental DIC data used by size and split (train, validation, test).}
\label{tab:dataset_splits}
\end{table}

After augmentation, the datasets comprise a total of 70,352 samples. The sample counts for each dataset size and data split are summarized in Table~\ref{tab:benchmark_dataset_sizes}.

\begin{table}
\begin{center}
\begin{tabular}{lccc}
\toprule
\textbf{Size} & \textbf{Train} & \textbf{Validation} & \textbf{Test} \\
\midrule
S & 10,048 & 5,944 & 5,944 \\
M & 21,640 & 11,736 & 11,672 \\
L & 42,056 & 11,736 & 16,560 \\
\bottomrule
\end{tabular}
\end{center}
\caption{Available Benchmark datasets and their sizes after the application of the augmentation strategy.}
\label{tab:benchmark_dataset_sizes}
\end{table}

\section*{Data Records}

\textbf{Repository and identifiers.} The dataset is archived on Zenodo (Version 2.0.0; published Feb 4, 2026) under the DOI \emph{10.5281/zenodo.18454958} with a CC BY 4.0 license \cite{CrackMNIST-Zenodo}. The record provides file-level MD5 checksums and a total hosted size of 21.1 GB.

\textbf{File inventory and structure.} Data are distributed as standalone HDF5 files named by pixel resolution and dataset size: \texttt{crackmnist\_\{28,64,128\}\_\{S, M, L\}.h5}. The Zenodo record hosts nine files for 28, 64, and 128 px (each with S, M, L). Each HDF5 file is self-contained; there is no nested directory structure besides the split organization described below.

\textbf{Contents and format.} Within each HDF5, samples are organized into three splits—\texttt{train}, \texttt{val}, and \texttt{test}. For every sample, the inputs are planar displacement fields $u_x, u_y$ (in mm) derived from digital image correlation, and the target is - depending on the chosen learning task - either a binary segmentation mask indicating the crack tip location or a vector containing $(K_I, K_{II}, T)$. The files are provided for three pixel resolutions ($28 \times 28, 64 \times 64, 128 \times 128$). The JSON file \texttt{experiments\_metadata.json} contains the experiment metadata, i.e. material and specimen types and thicknesses as well as fracture orientations and load ratios.

\section*{Technical Validation}
For technical validation of the dataset, benchmark studies were conducted for two tasks: crack tip detection (image-based inference) and stress intensity factor (SIF) prediction (physics-related regression). These two tasks represent core challenges in experimental fracture mechanics, namely pixel-precise crack tip localization and the prediction of global mechanical quantities characterizing the crack tip field.

The aim of these benchmarks is to provide baseline performance metrics across dataset sizes and resolutions. To ensure a clean and task-specific evaluation, a separate model was trained for each task.

\subsection*{Crack Tip Detection}

\textbf{Model architecture.}
For the crack tip detection task, we employed two convolutional neural networks (CNN) based on the U-Net architecture introduced by Ronneberger et al \cite{UNet}. The suitability of this architecture for crack tip detection was demonstrated in \cite{Strohmann21, Melching22}. The implementation was carried out in \texttt{PyTorch} \cite{PyTorch}. For the $28 \times 28$ px input patches, a shallow U-Net with two encoder and two decoder blocks was utilized, whereas for all higher resolutions the standard U-Net architecture consisting of four encoder and decoder blocks was employed. Both U-Net variants were initialized with 128 feature maps in the first convolutional block, two input channels ($u_x, u_y$), and a single output channel (crack tip class). A dropout rate of 0.2 was applied in the bottleneck layers. As in \cite{Strohmann21}, LeakyReLU \cite{LeakyReLU} was used as activation function. 
To ensure all input features ($u_x$, $u_y$) are on the same scale and crack tip detection is independent of the applied external load, instance normalization to mean 0 and standard deviation 1 was performed channel-wise.
The networks were then trained for 100 epochs using the Adam \cite{AdamOptimizer} optimizer in combination with a Dice loss function, acknowledging the extreme class imbalance between crack tip and background in crack detection. The training data was further divided into mini-batches of predefined batch sizes ($bs$). Random sampling was applied to the training sets. To provide initial training stabilization, a learning rate ($lr$) scheduler with a linear warm-up phase over 5 epochs followed by exponential learning rate decay ($\gamma$) was employed.

\textbf{Hyperparameter study.}
The study was performed on a on a Fujitsu Celsius R930 Power workstation with 2x 18-core Intel Xeon Gold 6240 3.9 GHz 25MB and a total of 512 GB RAM. For the training \& evaluation phase, a Nvidia Quatro RTX 8000 GPU was leveraged. 
Hyperparameter screening was conducted with \texttt{Optuna} \cite{optuna_2019} using the built-in Tree-structured Parzen Estimator (TPE) sampler. The optimization strategy chosen for \texttt{Optuna} was the minimization of the validation loss. Otherwise, default settings (see \texttt{Optuna} documentation) were found to be sufficient. The search was performed for $bs$, $lr$ and $\gamma$ as optimization parameters. $bs$ was explored in categorical values $\{8, 16, 32, 64\}$. $lr$ was sampled on a logarithmic scale between $10^{-5}$ and $10^{-2}$. $\gamma$ was allowed to vary between 0.9 and 1.0. The maximum training length was fixed to 100 epochs, with early pruning of under-performing trials allowed after epoch 20. Pruning was disabled for the first 10 trials to allow for a reliable performance baseline. A total of 50 trials (including pruned trials) were executed before optimization was stopped. The model parameters of the epoch with the lowest validation loss then define the final best model of each trial.
In light of the early trends observed during optimization, the hyperparameter optimization was limited to input resolutions $28 \times 28$, $64 \times 64$, as training for a comparable amount of trials at higher resolutions would have resulted in impractically total runtimes. Instead, the best hyperparameters of the $64 \times 64$ models were reused for the corresponding $128 \times 128$ models, as both share the same underlying U-Net architecture. Consequently, similar optimization behavior is expected.

Table~\ref{tab:best_hyperparameter_CTD} summarizes the best hyperparameter configurations identified for each dataset combination. For description, we use the shorthand notation \textit{Size-Resolution} (e.g., \textit{S-28}). Across most configurations, the largest batch size ($bs = 64$) emerged as the optimal choice, with the exception of the smallest setting (\textit{S-28}), which behaved as an outlier.
The learning rate ($lr$) consistently converged to values in the range $3 \times 10^{-3}$ to $5 \times 10^{-3}$ for most setups, with a trend towards slightly lower values ($<10^{-3}$) for the deeper, standard U-Net architecture and higher image resolutions. The decay parameter $\gamma$ was stable across configurations, typically falling between $0.93$ and $0.96$. Larger datasets tended to prefer slower decay (higher $\gamma$), whereas the deeper architecture favored faster decay (lower $\gamma$). 

\begin{table}[ht]
\centering
\begin{tabularx}{\linewidth}{c c *{3}{>{\centering\arraybackslash}X}}
\toprule
\multirow{2}{*}{\textbf{Size}} & \multirow{2}{*}{\textbf{Resolution}} & \multicolumn{3}{c}{\textbf{Best Hyperparameters}} \\
\cmidrule(lr){3-5}
 & & $bs$ & $lr$ [$\times 10^{-4}$] & $\gamma$ \\
\midrule
\multirow{2}{*}{S} & $28 \times 28$ & 8 & 34.753 & 0.930 \\
                   & $64 \times 64$ & 64 & 44.060 & 0.939 \\
\midrule
\multirow{2}{*}{M} & $28 \times 28$ & 64 & 35.639 & 0.944 \\
                   & $64 \times 64$ & 64 & 6.066 & 0.926 \\
\midrule
\multirow{2}{*}{L} & $28 \times 28$ & 64 & 52.866 & 0.958 \\
                   & $64 \times 64$ & 64 & 3.585 & 0.939 \\
\bottomrule
\end{tabularx}
\caption{Best hyperparameter settings identified for the crack tip detection task for each dataset size and input resolution. Shown are the optimal batch size ($bs$), learning rate ($lr$), and exponential learning rate decay factor ($\gamma$) as determined by \texttt{Optuna}’s hyperparameter search.}
\label{tab:best_hyperparameter_CTD}
\end{table}

Within the sampled parameter distribution, parameter importance was evaluated \textit{post hoc} using \texttt{Optuna}’s built-in fANOVA analysis~\cite{fANOVA} (see Table \ref{tab:importance_hyperparameter_CTD}). The analysis is therefore independent of the sampling strategy. Overall, the analysis revealed a clear hierarchy in parameter relevance: $bs$ showed the lowest importance, whereas the gain parameter $lr$ and $\gamma$ were rated higher on the importance rating. From a practical perspective, this supports maximizing the batch size (within hardware constraints) to reduce training time without compromising accuracy. Across the explored hyperparameter ranges, most configurations yielded comparable validation losses. Only combinations of very fast decay ($\gamma \rightarrow 0.9$) and very low learning rates ($lr \rightarrow 10^{-5}$) led to premature convergence into suboptimal local minima.

\begin{table}[ht]
\centering
\begin{tabularx}{\linewidth}{c c *{3}{>{\centering\arraybackslash}X}}
\toprule
\multirow{2}{*}{\textbf{Size}} & \multirow{2}{*}{\textbf{Resolution}} & \multicolumn{3}{c}{\textbf{fANOVA Hyperparameter importance}} \\
\cmidrule(lr){3-5}
 & & $bs$ & $lr$ & $\gamma$ \\
\midrule
\multirow{2}{*}{S} & $28 \times 28$ & 0.14 & \textbf{0.54} & 0.32 \\
                   & $64 \times 64$ & 0.23 & \textbf{0.63} & 0.14 \\
\midrule
\multirow{2}{*}{M} & $28 \times 28$ & 0.23 & \textbf{0.61} & 0.16 \\
                   & $64 \times 64$ & 0.18 & 0.20 & \textbf{0.62} \\
\midrule
\multirow{2}{*}{L} & $28 \times 28$ & 0.23 & 0.24 & \textbf{0.54} \\
                   & $64 \times 64$ & 0.09 & \textbf{0.46} & 0.45 \\
\bottomrule
\end{tabularx}
\caption{Crack tip segmentation task global hyperparameter importance computed by the fANOVA estimator for each size and resolution. Highest importance scores per size and resolution in bold.}
\label{tab:importance_hyperparameter_CTD}
\end{table}

\textbf{Training Benchmark.}
Table~\ref{tab:benchmark_CTD} summarizes the crack tip detection benchmark results for each size (S, M, L) and resolution (28 px, 64 px, 128 px) on training, validation, and test splits. For each configuration, the reported value represents the arithmetic mean of the average losses obtained over 10 independent training runs using the best-performing hyperparameter settings (see also Tab.~\ref{tab:best_hyperparameter_CTD}). To quantify variability, the standard deviation computed over these per-run mean losses is also provided.  
Across all dataset sizes, the deeper standard architecture consistently outperforms the shallower U-Net. For $28 \times 28$ inputs, the crack tip is represented by a single pixel, effectively reducing the segmentation to a binary classification at a single location. As a consequence, the loss becomes highly sensitive to minor localization errors. At higher input size, the crack tip mask spans a larger area ($64 \times 64$\,px $\rightarrow$ $3 \times 3$ foreground mask, $128 \times 128$\,px $\rightarrow$ $5 \times 5$ foreground mask). Imperfect but overlapping predictions are still in partial agreement with the ground truth. Consequently, the larger mask relaxes the localization requirement and can result in a lower loss values. For the $128 \times 128$ inputs, the models exhibits an increase in training, validation, and test losses compared to their corresponding $64 \times 64$ models, from which the training hyperparameters were inherited. The increase is likely due to the hyperparameters only being optimized for $64 \times 64$ and the higher task complexity associated with full spatial overlap between prediction and ground truth at $128 \times 128$. Moreover, although adapted by increasing the foreground mask from $3\times 3$ to $5\times 5$, class imbalance still slightly increases. 
For the S and L dataset sizes, the loss consistently increases from the training to the validation and test subsets, independent of image resolution. Dataset M is an exception, as its test loss is lower than the validation loss, which is attributed to statistical variability in the crack tip distributions across the splits. The largest relative increase from training to validation/test losses occurs for \textit{L-64}, with an approximately six-fold rise. 
Given the overall trend, these effect are attributable to a deliberate dataset design. In the case of Size L, the test subsets contain a unique sample type and material orientation (CT75\_7010\_SL45\_1\_right). This shift from training and validation data distribution was intentionally included to evaluate the model’s ability to generalize to unseen crack configurations and to provide a more realistic assessment of model robustness. 
Overall the standard deviation is low across all combinations for the crack tip detection task.

\begin{table}[ht]
\centering
\begin{tabularx}{\linewidth}{c c *{3}{>{\centering\arraybackslash}X}}
\toprule
\multirow{2}{*}{\textbf{Size}} & \multirow{2}{*}{\textbf{Resolution}} & \multicolumn{3}{c}{\textbf{Mean Loss $\pm$ Std}} \\
\cmidrule(lr){3-5}
 & & \textbf{Train} & \textbf{Val} & \textbf{Test} \\
\midrule
\multirow{3}{*}{S} & $28 \times 28$ & 0.120 $\pm$ 0.008 & 0.207 $\pm$ 0.003 & 0.272 $\pm$ 0.008 \\
                   & $64 \times 64$ & 0.039 $\pm$ 0.011 & 0.136 $\pm$ 0.003 & 0.166 $\pm$ 0.005 \\
                   & $128 \times 128$ & 0.069 $\pm$ 0.013 & 0.152 $\pm$ 0.005 & 0.176 $\pm$ 0.004 \\
\midrule
\multirow{3}{*}{M} & $28 \times 28$ & 0.121 $\pm$ 0.013 & 0.255 $\pm$ 0.005 & 0.225 $\pm$ 0.006 \\
                   & $64 \times 64$ & 0.052 $\pm$ 0.010 & 0.182 $\pm$ 0.003 & 0.142 $\pm$ 0.002 \\
                   & $128 \times 128$ & 0.072 $\pm$ 0.012 & 0.198 $\pm$ 0.003 & 0.171 $\pm$ 0.004 \\
\midrule
\multirow{3}{*}{L} & $28 \times 28$ & 0.143 $\pm$ 0.016 & 0.238 $\pm$ 0.003 & 0.315 $\pm$ 0.009 \\
                   & $64 \times 64$ & 0.043 $\pm$ 0.012 & 0.178 $\pm$ 0.002 & 0.256 $\pm$ 0.002 \\
                   & $128 \times 128$ & 0.058 $\pm$ 0.013 & 0.199 $\pm$ 0.002 & 0.293 $\pm$ 0.002 \\
\bottomrule
\end{tabularx}
\caption{Crack tip detection benchmark results across 10 complete training runs using the best-performing hyperparameter configuration (see Table \ref{tab:best_hyperparameter_CTD})}
\label{tab:benchmark_CTD}
\end{table}

\textbf{Failure modes.}
Spatial loss maps were analyzed to identify characteristic failure modes. The losses are decomposed into false negatives (FN; GT=1) and false positives (FP; GT=0). To account for the non-uniform spatial distribution of crack tips, FN and FP maps are further normalized by the pixel-wise absolute crack tip occurrence. This normalization highlights relative error intensity (failure modes) in sparsely populated regions, otherwise overshadowed by statistical errors. It enables a comparison that is independent of the absolute crack-tip frequency. 

\begin{figure}
    \centering
    \includegraphics[width=0.9\linewidth]{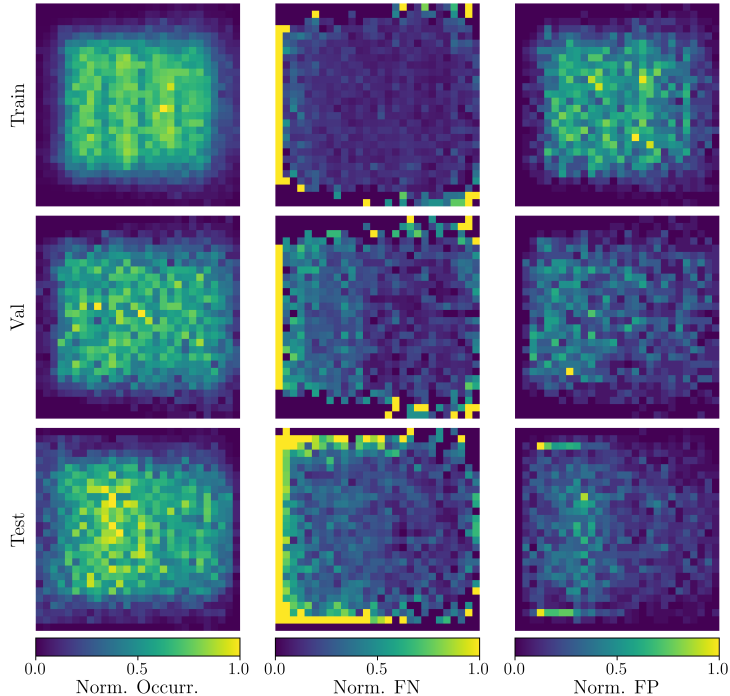}
    \caption{Prediction failure map for \textit{L-28}. Within each split, the mosaic shows crack tip occurrence (mask), false negative (FN) pixel and false positive (FP) pixel maps. The maps are normalized over per-pixel absolute crack tip occurrences and over trials and to highlight the spacial density distribution of correctly and incorrectly detected crack tips.}
    \label{fig:loss_summary}
\end{figure}

Figure~\ref{fig:loss_summary} shows prediction failure maps for resolution $28 \times 28$ and dataset size L. Each tile corresponds to a normalized spatial heat map. Within each split, the first column shows the normalized ground truth (GT) mask as crack tip occurrence, the second column as normalized FN and the third column as normalized FP maps. A consistent spatial error pattern emerges. Crack tips located near image boundaries and at the edges of the spatial distribution represented in the training data are less reliably detected. This indicates that the model is biased towards the crack-tip distribution present during training. Consequently, crack tips occurring in initially underrepresented or rarely observed spatial regions are more prone to FN predictions. In contrast, FP predictions appear more uniformly, which suggests small spatial misalignments between predicted and target masks rather than systematic hallucination.

\subsection*{SIF regression}
\textbf{Model architecture.}
For technical validation of the SIF regression task, we employ a convolutional encoder--regressor architecture. The encoder corresponds to the feature-extraction stage of the U-Net models used previously for the crack tip segmentation. Its bottleneck output is global-average pooled and passed to a lightweight multi-layer perceptron (MLP). The MLP comprises a single hidden layer with 256 units, ReLU activation, and dropout ($p=0.2$), followed by a linear output layer mapping the 256-dimensional representation to the three target crack tip field descriptors $K_I$, $K_{II}$, and $T$. This deliberately compact CNN+MLP architecture provides a clean and interpretable baseline for benchmarking the regression task.

In contrast to the segmentation setup, the input displacement fields $(u_x, u_y)$ are not normalized, as their magnitudes are directly linked to the applied load and therefore carry information predictive of the SIFs. Instead, each descriptor (target) is standardized, using its mean and standard deviation (\textit{z}-score) computed from the training subset. Across the full dataset, $K_I$ ranges approximately from almost $0$ to $40~\text{MPa}\sqrt{\text{m}}$, $K_{II}$ from $-3$ to $2~\text{MPa}\sqrt{\text{m}}$, and $T$ from $-100$ to $50~\text{MPa}$.

\textbf{Hyperparameter study.}
Hyperparameter screening was conducted following the same procedure as for the crack tip segmentation task. The number of trials was reduced to 25 without measurable loss in optimization quality. Table~\ref{tab:best_hyperparameter_SIF} summarizes the optimal hyperparameter configurations for the SIF regression task across all dataset sizes and resolutions.
Overall, the results are consistent with those obtained for crack tip segmentation. Larger batch sizes were generally advantageous. Compared to the the crack tip detection task, the optimal learning rates are approximately one to two orders of magnitude lower. Based on these findings, we recommend maximizing the batch size within the hardware constraints to reduce training time.

\begin{table}[ht]
\centering
\begin{tabularx}{\linewidth}{c c *{3}{>{\centering\arraybackslash}X}}
\toprule
\multirow{2}{*}{\textbf{Size}} & \multirow{2}{*}{\textbf{Resolution}} & \multicolumn{3}{c}{\textbf{Best Hyperparameters}} \\
\cmidrule(lr){3-5}
 & & $bs$ & $lr$ [$\times 10^{-4}$] & $\gamma$ \\
\midrule
\multirow{2}{*}{S} & $28 \times 28$ & 64 & 3.431 & 0.976 \\
                   & $64 \times 64$ & 32 & 0.861 & 0.952 \\
\midrule
\multirow{2}{*}{M} & $28 \times 28$ & 32 & 7.470 & 0.904 \\
                   & $64 \times 64$ & 64 & 15.515 & 0.907 \\
\midrule
\multirow{2}{*}{L} & $28 \times 28$ & 64 & 13.985 & 0.930 \\
                   & $64 \times 64$ & 64 & 4.484 & 0.970 \\
\bottomrule
\end{tabularx}
\caption{Best hyperparameter settings identified for the SIF regression task for each dataset size and input resolution. Shown are the optimal batch size ($bs$), learning rate ($lr$), and exponential learning rate decay factor ($\gamma$) as determined by \texttt{Optuna}’s hyperparameter search.}
\label{tab:best_hyperparameter_SIF}
\end{table}

In contrast to the crack tip detection task, the highest ranked hyperparameter importance for the SIF regression task is more evenly distributed across batch size, $bs$, learning rate, $lr$, and decay rate, $\gamma$ (see Table \ref{tab:importance_hyperparameter_SIF}). However, this observation should be interpreted with caution, as the \texttt{Optuna} study was conducted with a reduced number of trials. Consequently, the corresponding \textit{fANOVA}-based importance estimates are expected to be less reliable. Nevertheless, no single hyperparameter emerged as a dominant factor within the explored search space.

\begin{table}[ht]
\centering
\begin{tabularx}{\linewidth}{c c *{3}{>{\centering\arraybackslash}X}}
\toprule
\multirow{2}{*}{\textbf{Size}} & \multirow{2}{*}{\textbf{Resolution}} & \multicolumn{3}{c}{\textbf{fANOVA Hyperparameter importance}} \\
\cmidrule(lr){3-5}
 & & $bs$ & $lr$ & $\gamma$ \\
\midrule
\multirow{2}{*}{S} & $28 \times 28$ & 0.17 & \textbf{0.59} & 0.24 \\
                   & $64 \times 64$ & 0.07 & 0.19 & \textbf{0.74} \\
\midrule
\multirow{2}{*}{M} & $28 \times 28$ & \textbf{0.26} & 0.14 & 0.6 \\
                   & $64 \times 64$ & 0.30 & \textbf{0.52} & 0.18 \\
\midrule
\multirow{2}{*}{L} & $28 \times 28$ & \textbf{0.64} & 0.21 & 0.16 \\
                   & $64 \times 64$ & 0.11 & \textbf{0.54} & 0.35 \\
\bottomrule
\end{tabularx}
\caption{SIF regression task global hyperparameter importance computed by the fANOVA estimator for each size and resolution. Highest importance scores per size and resolution in bold.}
\label{tab:importance_hyperparameter_SIF}
\end{table}

\textbf{Training Benchmark.}
Table~\ref{tab:benchmark_SIF} summarizes the benchmark results for SIF regression across dataset sizes and input resolutions. Overall, all models successfully fit the training data but exhibit pronounced overfitting. For the $28 \times 28$ resolution, the training loss is approximately 20-40 \% of the validation loss. For $64 \times 64$ and $128 \times 128$ inputs, overfitting is less pronounced, with training losses approximately 15-20\% of the corresponding validation loss.
Across all dataset sizes, model performance generally improves with increasing input resolution, although this effect is less pronounced than in the crack tip detection task. In some size-–resolution-–split combinations, the deeper U-Net architecture exhibits a higher validation loss than the shallow architecture at $28 \times 28$ (e.g. val--\textit{S-28}). Furthermore, the tendency towards higher losses at $128 \times 128$ compared to $64 \times 64$ is less evident than in the detection task. Loss levels across dataset sizes are comparable when evaluated at the same resolution.
Compared to the crack tip detection, the SIF regression task exhibits a higher relative variability over the repeated runs, with the standard deviation reaching approximately 10 \% of the corresponding mean loss values. The increased variability across repeated runs further suggests a more complex and less stable training for SIF regression compared to crack tip segmentation.

\begin{table}[ht]
\centering
\begin{tabularx}{\linewidth}{c c *{3}{>{\centering\arraybackslash}X}}
\toprule
\multirow{2}{*}{\textbf{Size}} & \multirow{2}{*}{\textbf{Resolution}} & \multicolumn{3}{c}{\textbf{Mean Loss $\pm$ Std}} \\
\cmidrule(lr){3-5}
 & & \textbf{Train} & \textbf{Val} & \textbf{Test} \\
\midrule
\multirow{2}{*}{S} & $28 \times 28$ & 0.068 $\pm$ 0.048 & 0.183 $\pm$ 0.009 & 0.462 $\pm$ 0.048 \\
                   & $64 \times 64$ &  0.019 $\pm$ 0.026 & 0.217 $\pm$ 0.018 & 0.447 $\pm$ 0.048 \\
                   & $128 \times 128$ & 0.020 $\pm$ 0.024 & 0.195 $\pm$ 0.018 & 0.364 $\pm$ 0.055 \\
\midrule
\multirow{2}{*}{M} & $28 \times 28$ & 0.079 $\pm$ 0.025 & 0.189 $\pm$ 0.008 & 0.402 $\pm$ 0.032 \\
                   & $64 \times 64$ &  0.020 $\pm$ 0.008 & 0.181 $\pm$ 0.019 & 0.366 $\pm$ 0.044 \\
                   & $128 \times 128$ & 0.025 $\pm$ 0.010 & 0.184 $\pm$ 0.007 & 0.449 $\pm$ 0.028 \\
\midrule
\multirow{2}{*}{L} & $28 \times 28$ & 0.034 $\pm$ 0.012 & 0.161 $\pm$ 0.004 & 0.409 $\pm$ 0.028 \\
                   & $64 \times 64$ & 0.020 $\pm$ 0.012 & 0.134 $\pm$ 0.005 & 0.365 $\pm$ 0.019 \\
                   & $128 \times 128$ & 0.019 $\pm$ 0.012 & 0.134 $\pm$ 0.011 & 0.350 $\pm$ 0.019 \\
\bottomrule
\end{tabularx}
\caption{SIF regression benchmark results across 10 complete training runs using the best-performing hyperparameter configuration (see Table \ref{tab:best_hyperparameter_SIF})}
\label{tab:benchmark_SIF}
\end{table}

\begin{figure}[ht]
    \centering
    \includegraphics[width=1.0\linewidth]{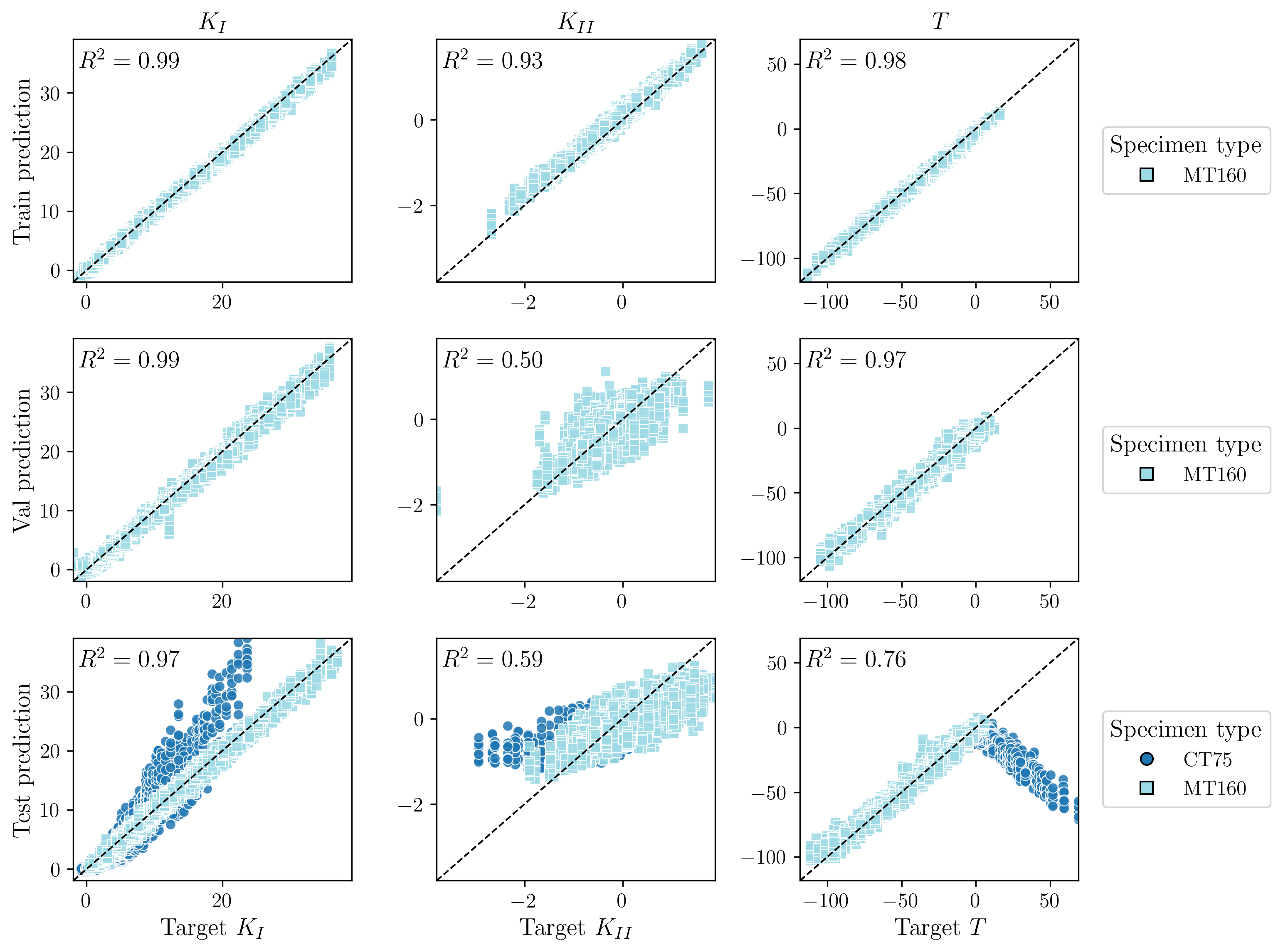}
    \caption{Predicted versus target $K_I$, $K_{II}$, and $T$ for the best individual SIF regression model of the \textit{L-64} training benchmark. Each training, validation and test split subplot reports the coefficient of determination $R^2$. The dashed line denotes perfect agreement. Each data point is also categorised by specimen type.}
    \label{fig:r2_L64px}
\end{figure}

\textbf{Failure modes.}
In order to gain deeper insight into the models failure modes, Figure~\ref{fig:r2_L64px} presents a detailed comparison of predicted and target values of $K_I$, $K_{II}$, and $T$ for the best-performing SIF regression model from the representative \textit{L-64} benchmark. Results are shown separately for the training, validation, and test splits. 
Test dataset size L contains experiment CT75\_7010\_SL45\_1\_right, whose geometry and constraint deviates significantly from the MT160 configurations present in the training and validation split. As a consequence, the corresponding $T$ values are positive, in contrast to the negative $T$-stresses observed for the MT specimens. The model consistently fails to extrapolate to these positive $T$-stresses, which is expected for a standard MLP-based regressor trying to extrapolate beyond the confines of the training data distribution \cite{Barnard92, barrett2018}. This lack of extrapolation capabilities is also reflected in the lower $R^2$ for $K_I$, albeit less prominently. The figure shows, that the prediction error scales with $K_I$. This result indicates that the model learned a pattern akin to a weight function, strongly correlated with the MT160 configuration.
Detailed tables reporting the $R^2$ performance for each target variable $K_I, K_{II}$, and $T$, across all dataset sizes and input resolutions, are provided in the appendix. 

\section*{Usage Notes}
\textit{CrackMNIST} can be accessed either (i) via the accompanying Python package \texttt{crackmnist} \cite{CrackMNIST-GitHub} which downloads the requested dataset file fully automatically on demand, or (ii) by manually downloading the released HDF5 files and placing them into a local dataset root directory.

By default, the package uses a user cache directory (typically \texttt{\textasciitilde/.crackmnist} on Linux) as the dataset root. For each configuration, the package expects the corresponding HDF5 file \\ \texttt{crackmnist\_\{pixels\}\_\{size\}.h5}, where $\texttt{pixels} \in \{28,64,128\}$ and $\texttt{size}\in \{"S","M","L"\}$, and the experiment metadata file \texttt{experiments\_metadata.json} to be present in the dataset root.

\textbf{Recommended data loading patterns.}
The dataset is implemented as a \texttt{PyTorch} \texttt{Dataset}. A minimal usage pattern is:

\begin{lstlisting}[language=Python,basicstyle=\ttfamily\small,numbers=left]
from crackmnist import CrackMNIST

# Crack tip segmentation (binary mask target)
data_ct_seg = CrackMNIST(split="train", size="S", pixels=28, task="crack_tip_segmentation")

# SIF regression (K_I, K_II, T target)
data_sif_reg = CrackMNIST(split="train", size="S", pixels=28, task="SIF_regression")

x, y = ds_seg[0]   # x: (2, H, W), y: mask
x, y = ds_sif[0]   # x: (2, H, W), y: (3,) = (K_I, K_II, T)
\end{lstlisting}

Inputs are two-channel displacement fields $(u_x,u_y)$ measured in millimetres. Targets depend on the chosen \texttt{task}: a binary crack-tip segmentation mask for \texttt{crack\_tip\_segmentation}, or a three-component vector $(K_I, K_{II}, T)$ for \texttt{SIF\_regression} (with $K_I$ and $K_{II}$ in MPa$\sqrt{\mathrm{m}}$ and $T$ in MPa).

\textbf{File format and dataset keys.}
The released data are stored in HDF5. Each file contains split-specific arrays for inputs and targets:
\begin{itemize}
\item \texttt{\{split\}\_images}: displacement fields (two channels).
\item \texttt{\{split\}\_masks}: binary crack-tip masks (segmentation task).
\item \texttt{\{split\}\_SIFs}: $(K_I, K_{II}, T)$ targets (regression task).
\item \texttt{\{split\}\_exp\_ids}: experiment identifiers for each sample.
\item \texttt{experiments}: experiment name strings indexed by \texttt{exp\_ids}.
\item \texttt{\{split\}\_forces}: scalar load/force value per sample (see accompanying metadata for definition).
\item \texttt{\{split\}\_augs}: augmentation parameters applied to each sample.
\end{itemize}
The companion JSON file \texttt{experiments\_metadata.json} stores per-experiment metadata and can be queried through the dataset helper \texttt{get\_metadata(idx)}. The functions \texttt{get\_forces(idx)} and \texttt{get\_augmentations(idx)} provide convenient access to per-sample forces and augmentation parameters.

\textbf{Augmentations and transforms.}
For each sample, the augmentation record contains shift, rotation, and a vertical flip flag. Specifically, \texttt{get\_augmentations(idx)} returns a dictionary with
\texttt{shift=(shift\_x, shift\_y)} in millimetres, \texttt{rotation} in degrees, and \texttt{vertical\_flip} as a boolean indicator. These values can be used to reproduce or stratify analyses with respect to the applied geometric perturbations.

\begin{lstlisting}[language=Python,basicstyle=\ttfamily\small,numbers=left]
idx = 0  # Choose an index

force = data_ct_seg.get_forces(idx)  # applied force in N
metadata = data_ct_seg.get_metadata(idx)
augmentation = data_ct_seg.get_augmentations(idx)

print(f"Experiment: {metadata['experiment']}")
print(f"Material:   {metadata['material']}")
print(f"Thickness:  {metadata['thickness_mm']} mm")
print(f"Shift:      ({augmentation['shift'][0]:.2f}, {augmentation['shift'][1]:.2f}) mm")
print(f"Rotation:   {augmentation['rotation']:.2f}")
print(f"Flip (Y/N): {augmentation['vertical_flip']}")
\end{lstlisting}

The package optionally supports input/target transforms via \texttt{transform} and \texttt{target\_transform}. The provided \texttt{InputNormalization} transform performs per-sample, per-channel standardization (i.e., normalization parameters are computed from each sample unless fixed means/stds are provided).

\section*{Code Availability}
\textit{CrackMNIST} is published as an open-source Python package (\url{https://pypi.org/project/crackmnist/}). The codebase is available on GitHub (\url{https://github.com/dlr-wf/crackmnist}). The data annotation algorithm is published and implemented in \texttt{CrackPy} (\url{https://github.com/dlr-wf/crackpy}).

\section*{Acknowledgements}
The authors acknowledge the financial support of the German federal ministry of Economic Affairs and Climate Action (BMWK) within the LuFo-VI project "Untersuchung der Prozess-Struktur-Eigenschaftsbeziehungen rollgeformter und rollprofilierter Profile für eine kosten- und ökooptimierte Türumgebungsstruktur" (Funding ID 20W2102C). The research is also funded by the Deutsche Forschungsgemeinschaft (DFG, German Research Foundation)-Project number BR 6259/2-2. The authors acknowledge the financial support of the DLR Aeronautics Directorate.
We acknowledge the financial support of the DLR-Directorate Aeronautics.

\section*{Author contributions}
D.M. conceived the main idea of the publication, designed the datasets, curated and annotated the data, and wrote the manuscript. F.D. designed and conducted the FCG experiments, performed the ML baseline training and wrote the manuscript. F.P. designed and conducted the FCG experiments. E.S. implemented the \textit{CrackMNIST} Python software package. E.D. supported the implementation of the infrastructure and the FCG experiments. E.B. designed, managed, and supervised the FCG experimental work. All authors reviewed and commented on the manuscript.

\section*{Competing interests}
The Authors declare no competing financial or non-financial interests.

\printbibliography

\appendix

\section*{Appendix}
$K_I$, $K_{II}$, and $T$-stress represent distinct crack tip field descriptors and may therefore exhibit different sensitivities to model error. 
The coefficient of determination, $R^2$, between predicted and target values $K_I$, $K_{II}$, and $T$ accross all dataset sizes and input resolutions can be found in Table~\ref{tab:r2_KI}-\ref{tab:r2_T}. The reported values correspond to the mean and standard deviation of $R^2$ obtained from the ten independent benchmark training runs with the best hyperparameters. 
Across all three targets, the model tends to overfitting, with this effect being most pronounced for $K_{II}$. The dataset is derived from experiments conducted under nominal Mode I ($K_I$) loading conditions. Consequently, Mode II ($K_{II}$) is expected to be minimal. The rather small $K_{II}$ (-2 to 2 $MPa\sqrt{m}$) are attributable to secondary factors such as microstructure-induced crack kinking and experimental noise. As a result, $R^2$ values for $K_{II}$ on the validation and test data set are less than half of those observed for the training dataset, indicating poor generalization. This suggests that the model is unable to resolve $K_{II}$ in a physically meaningful way, as the $K_{II}$ signal might be dominated by noise rather than systematic effects.
Nevertheless, for $K_{II}$ a trend towards higher model performance appears at higher input resolution, indicating that information patterns related to Mode II SIFs may only become discernible at higher resolution. This behaviour is not observed for $K_I$ or $T$. Consequently, the lower mean losses for the training split can be mostly attributed to the better performance for $K_{II}$. 

The models demonstrated strong predictive performance for $K_I$ across all data sets and input resolutions. The lowest means $R^2=0.97$ is observed for size S in the validation split and size L in the test split, independent of resolution. Performance for $T$ is on a similar level, although larger deviations from the training $R^2$'s are observed. For the test split, Mean $R^2 \leq 0.9$ were observed for dataset sizes S and L, with size L showing the most pronounced degradation. 

\begin{table}[ht]
\centering
\begin{tabularx}{\linewidth}{c c *{3}{>{\centering\arraybackslash}X}}
\toprule
\multirow{2}{*}{\textbf{Size}} & \multirow{2}{*}{\textbf{Resolution}} 
& \multicolumn{3}{c}{\textbf{$R^2$ ($K_I$)}} \\
\cmidrule(lr){3-5}
 & & \textbf{Train} & \textbf{Val} & \textbf{Test} \\
\midrule
\multirow{3}{*}{S} 
 & $28 \times 28$ & 0.99 $\pm$ 0.01 & 0.97 $\pm$ 0.01 & 0.98 $\pm$ 0.01 \\
 & $64 \times 64$ & 0.99 $\pm$ 0.00 & 0.97 $\pm$ 0.01 & 0.98 $\pm$ 0.00 \\
 & $128 \times 128$ & 0.99 $\pm$ 0.01 & 0.98 $\pm$ 0.00 & 0.98 $\pm$ 0.00 \\
\midrule
\multirow{3}{*}{M} 
 & $28 \times 28$ & 0.99 $\pm$ 0.00 & 0.99 $\pm$ 0.00 & 0.98 $\pm$ 0.00 \\
 & $64 \times 64$ & 0.99 $\pm$ 0.00 & 0.99 $\pm$ 0.00 & 0.99 $\pm$ 0.00 \\
 & $128 \times 128$  & 0.99 $\pm$ 0.00 & 0.99 $\pm$ 0.00 & 0.99 $\pm$ 0.00 \\
\midrule
\multirow{3}{*}{L} 
 & $28 \times 28$ & 0.99 $\pm$ 0.00 & 0.99 $\pm$ 0.00 & 0.97 $\pm$ 0.01 \\
 & $64 \times 64$ & 1.00 $\pm$ 0.00 & 0.99 $\pm$ 0.00 & 0.97 $\pm$ 0.01 \\
 & $128 \times 128$  & 1.00 $\pm$ 0.00 & 0.99 $\pm$ 0.00 & 0.97 $\pm$ 0.00 \\
\bottomrule
\end{tabularx}
\caption{$R^2$ performance for mode I stress intensity factor $K_I$ across dataset sizes and resolutions. Means and standard deviation of the 10 training runs.}
\label{tab:r2_KI}
\end{table}

\begin{table}[ht]
\centering
\begin{tabularx}{\linewidth}{c c *{3}{>{\centering\arraybackslash}X}}
\toprule
\multirow{2}{*}{\textbf{Size}} & \multirow{2}{*}{\textbf{Resolution}} 
& \multicolumn{3}{c}{\textbf{$R^2$ ($K_{II}$)}} \\
\cmidrule(lr){3-5}
 & & \textbf{Train} & \textbf{Val} & \textbf{Test} \\
\midrule
\multirow{3}{*}{S} 
 & $28 \times 28$  & 0.84 $\pm$ 0.10 & $-0.16 \pm 0.12$ & $-0.00 \pm 0.09$ \\
 & $64 \times 64$ & 0.96 $\pm$ 0.06 & $-0.38 \pm 0.19$ & 0.07 $\pm$ 0.08 \\
 & $128 \times 128$  & 0.96 $\pm$ 0.06 &$-0.28 \pm 0.14$ & 0.23 $\pm$ 0.12 \\
\midrule
\multirow{3}{*}{M} 
 & $28 \times 28$ & 0.83 $\pm$ 0.07 & 0.41 $\pm$ 0.03 & 0.13 $\pm$ 0.07 \\
 & $64 \times 64$ & 0.95 $\pm$ 0.02 & 0.45 $\pm$ 0.05 & 0.23 $\pm$ 0.09 \\
 & $128 \times 128$  & 0.95 $\pm$ 0.02 & $0.43 \pm 0.03$ & 0.03 $\pm$ 0.07 \\
\midrule
\multirow{3}{*}{L} 
 & $28 \times 28$ & 0.92 $\pm$ 0.03 & 0.40 $\pm$ 0.02 & 0.33 $\pm$ 0.05 \\
 & $64 \times 64$ & 0.96 $\pm$ 0.03 & 0.51 $\pm$ 0.02 & 0.54 $\pm$ 0.03 \\
 & $128 \times 128$  & 0.95 $\pm$ 0.03 & $0.50 \pm 0.05$ & 0.45 $\pm$ 0.05 \\
\bottomrule
\end{tabularx}
\caption{$R^2$ performance for mode II stress intensity factor $K_{II}$ across dataset sizes and resolutions. Means and standard deviation of the 10 training runs.}
\label{tab:r2_KII}
\end{table}

\begin{table}[ht]
\centering
\begin{tabularx}{\linewidth}{c c *{3}{>{\centering\arraybackslash}X}}
\toprule
\multirow{2}{*}{\textbf{Size}} & \multirow{2}{*}{\textbf{Resolution}} 
& \multicolumn{3}{c}{\textbf{$R^2$ ($T$)}} \\
\cmidrule(lr){3-5}
 & & \textbf{Train} & \textbf{Val} & \textbf{Test} \\
\midrule
\multirow{3}{*}{S} 
& $28 \times 28$ & 0.97 $\pm$ 0.04 & 0.94 $\pm$ 0.02 & 0.90 $\pm$ 0.02 \\
 & $64 \times 64$ & 0.99 $\pm$ 0.01 & 0.94 $\pm$ 0.03 & 0.87 $\pm$ 0.03 \\
 & $128 \times 128$ & 0.99 $\pm$ 0.01 & 0.95 $\pm$ 0.01 & 0.90 $\pm$ 0.02 \\
\midrule
\multirow{3}{*}{M} 
& $28 \times 28$ & 0.98 $\pm$ 0.01 & 0.97 $\pm$ 0.01 & 0.95 $\pm$ 0.01 \\
 & $64 \times 64$ & 0.99 $\pm$ 0.00 & 0.97 $\pm$ 0.01 & 0.96 $\pm$ 0.01 \\
  & $128 \times 128$ & 0.99 $\pm$ 0.01 & 0.97 $\pm$ 0.00 & 0.96 $\pm$ 0.01 \\
\midrule
\multirow{3}{*}{L} 
 & $28 \times 28$ & 0.99 $\pm$ 0.01 & 0.98 $\pm$ 0.01 & 0.83 $\pm$ 0.03 \\
 & $64 \times 64$ & 0.99 $\pm$ 0.01 & 0.98 $\pm$ 0.01 & 0.79 $\pm$ 0.02 \\
 & $128 \times 128$ & 0.99 $\pm$ 0.01 & 0.95 $\pm$ 0.01 & 0.84 $\pm$ 0.02 \\
\bottomrule
\end{tabularx}
\caption{$R^2$ performance for T stress $T$ across dataset sizes and resolutions. Means and standard deviation of the 10 training runs.}
\label{tab:r2_T}
\end{table}

\end{document}